\documentclass[twocolumn,showpacs,preprintnumbers,amsmath,amssymb,floatfix]{revtex4}

\usepackage{epsfig,psfrag}
\usepackage{dcolumn}
\usepackage{bm}
\usepackage{graphicx}
\usepackage{color}

\begin{document}

\title{Josephson-current induced conformational switching of a
 molecular quantum dot}
\author{A. Zazunov, A. Schulz, and R. Egger}
\affiliation{Institut f\"ur 
Theoretische Physik, Heinrich-Heine-Universit\"at,
D-40225  D\"usseldorf, Germany}
\date{\today}

\begin{abstract}
We discuss the behavior of a two-level system coupled to a quantum 
dot contacted by superconducting source/drain electrodes, 
representing a simple model for the conformational degree
of freedom of a molecular dot or a break junction.  The Josephson
current is shown to induce conformational changes, including 
a complete reversal.  For small bias voltage, periodic 
conformational motions induced by Landau-Zener transitions
between Andreev states are predicted. 
\end{abstract}

\maketitle

The remarkable recent progress in the fabrication and experimental
study of transport through ultrasmall nanoscopic devices,
break junctions, or molecules
(in the following termed ``quantum dot'' or simply ``dot'') \cite{molel}
has stimulated renewed interest in the Josephson effect \cite{golubov},
where the Josephson current through a dot contacted 
by superconducting electrodes with phase difference $\varphi$ 
is the relevant observable.  The full current-phase relation has been
 measured in various systems, and electron-electron interactions
on the dot were shown to be important \cite{CPR-exp}, as expected from
theory \cite{bcs-anderson}.  The already achieved wide tunability 
(via gate electrodes) and impressive control over Josephson currents 
through nanoscale dots indicate that experiments should be able to also
probe modifications of the super-current due to the coupling of the dot 
to another quantum system (e.g., a spin or a side-coupled dot).
Many previous efforts have focussed on studying the coupling to the spin degree 
of freedom in molecular magnets \cite{molmag}, which is also related to issues appearing for 
superconductor-ferromagnet-superconductor structures \cite{SFS}.
Theoretical work has also discussed the effects of local vibration modes 
on the super-current, where the dot is coupled to a 
boson mode (phonon) \cite{bcs-holstein1,bcs-holstein2}.

Surprisingly, so far the effects of a {\sl two-level system} (TLS) coupled
to the dot have not been addressed, except for  normal-conducting
leads. This is an important question, since for instance
two {\sl conformational configurations} of a molecule may
represent the TLS degree of freedom. Experimental results  
for molecular dots or break junctions
(with normal leads) were interpreted along this line 
\cite{tls-exp1,tls-exp2,tls-the,fabrizio}, but a TLS can also be realized
for a side-coupled double-dot system in the Coulomb 
blockade regime \cite{foot0}.  For concreteness,
we here refer to the TLS states $\sigma_z=\pm 1$ 
as the two distinct conformational states of a molecular dot,
where $\sigma_z$ couples to the dot's {\sl charge}.
A coupling of the TLS to the dot's spin does not have
a significant effect on the phenomena of interest here, see below.
(In any case, spin effects have been addressed in different 
contexts previously \cite{molmag}.) Our theory indicates 
that by variation of the phase $\varphi$, the TLS state can be 
significantly affected over a wide parameter regime, including a 
complete reversal of the conformational configuration.  
This remarkable effect allows for the dissipationless control (including
 switching) of the conformational degree of freedom ($\sigma_z$) 
in terms of the phase difference $\varphi$, which
can be tuned experimentally by embedding the device 
in a SQUID geometry \cite{CPR-exp}.  Conversely, changing the
conformational state will affect the Josephson current in a distinct
manner.  Moreover, when applying a bias voltage, a periodic
conformational motion is triggered via the ac Josephson effect 
involving Landau-Zener (LZ) transitions between Andreev states.
Our predictions (both for zero and finite bias) can be tested experimentally 
for a wide class of molecules electrically contacted in a break junction setup. 
Related experiments, reporting TLS behavior due to a conformational 
variable, have been published for normal leads \cite{tls-exp1}.
For normal leads, the model employed below has also been 
motivated in a recent theoretical work \cite{fabrizio}.
Available parameter estimates for dot and TLS energy scales
\cite{tls-exp1,fabrizio} suggest that the
predicted phenomena can be observed using existing state-of-the-art
experiments. Detection schemes to read out the conformational state are 
also available, e.g.  by single-molecule force microscopy \cite{zhang}.

We study a spin-degenerate molecular dot level with single-particle
energy $\epsilon_d$ and on-site Coulomb repulsion $U>0$, 
coupled to the TLS and to two superconducting banks (leads).  
Employing the standard wide-band approximation for the leads,
we assume a symmetric situation \cite{foot}, where
the banks are modelled as identical $s$-wave BCS superconductors with
gap $\Delta$ and the dot-lead hybridizations are equal, $\Gamma_L=
\Gamma_R= \Gamma/2$.  The TLS describing the conformational state
corresponds to Pauli matrices $\sigma_{x,z}$,
with bare energy difference $E_0$ and tunnel matrix element $W_0$ between
the two states.  The Hamiltonian is $H=H_0 + H_{\rm tun}
+ H_{\rm leads}$, where the coupled dot-plus-TLS part is 
(we set $e=\hbar=k_B=1$) 
\begin{equation}\label{h0}
H_0 = -\frac{E_0}{2} \sigma_z - \frac{W_0}{2} \sigma_x  
+ \left( 
\epsilon_d + \frac{\lambda}{2} \sigma_z \right) (n_\uparrow +n_\downarrow)
+ U n_\uparrow n_\downarrow,
\end{equation}
with the  occupation number $n_s= d^\dagger_s d^{}_s$
for the dot fermion $d_s$ with spin $s=\uparrow,\downarrow$,
$H_{\rm leads}$ describes standard BCS Hamiltonians,
$\varphi$ can be included by phase factors in the tunnel 
Hamiltonian $H_{\rm tun}$ \cite{bcs-holstein2},
and we define the renormalized dot level $\epsilon=\epsilon_d+U/2.$
In Eq.~\eqref{h0}, the TLS couples with strength $\lambda$ to the 
charge on the dot, which can be rationalized in simple terms by assuming
a one-dimensional effective reaction coordinate $X$ describing 
conformations of the molecule.  The dominant coupling to the 
electronic degrees of freedom is then (as for phonons) of the
 form $\propto X(n_\uparrow+ n_\downarrow)$ 
\cite{bcs-holstein1,bcs-holstein2,tls-exp1,tls-exp2}.
In the limit of interest, the potential energy $V(X)$ is bistable 
with two local minima, and a truncation of the low-energy dynamics of
$X$ to the lowest quantum state in each well leads to Eq.~\eqref{h0}.  
For a detailed derivation, see also Ref.~\cite{fabrizio}.

When dealing with the equilibrium problem, it is convenient  
to work with Nambu spinors $d(\tau)=(d_\uparrow, d^\dagger_\downarrow)$
in imaginary time $\tau.$ The lead fermions can then be integrated out exactly, 
and the partition function is 
$Z= {\rm Tr} \left( e^{-H_0/T} {\cal T} e^{-\int d\tau d\tau'
d^\dagger(\tau) \Sigma(\tau-\tau') d(\tau')} \right),$
where the trace extends over the dot-plus-TLS degrees of freedom only, 
${\cal T}$ denotes time-ordering, and the effect of the BCS leads
is contained in the $2\times 2$ Nambu self-energy matrix $\Sigma(\tau)$,
whose Fourier transform is \cite{bcs-anderson} 
\begin{equation}\label{sigma}
\Sigma(\omega;\varphi) =   \frac{\Gamma}{\sqrt{\omega^2+\Delta^2}}
\left(  \begin{array}{cc}- i\omega & 
\Delta \cos(\varphi/2) \\
\Delta \cos(\varphi/2) & - i\omega \end{array} \right).
\end{equation}
We mostly consider zero temperature, $T=0$,  where
both the Josephson current $I(\varphi)$ through the dot
and the expectation value $S=\langle \sigma_z \rangle$ of the 
conformational state follow from the ground-state energy
$E_g(\varphi, E_0)$ according to 
\begin{equation}\label{current}
I(\varphi) = 2 \frac{\partial E_g}{\partial \varphi},
\quad S(\varphi) = -2 \frac{\partial E_g}{\partial E_0} .
\end{equation}
Later on the formalism will be extended along the lines
of Refs.~\cite{averin,alfredo} to allow for the description
of a small bias voltage $V$ as well.

\begin{figure}
\scalebox{0.32}{\includegraphics{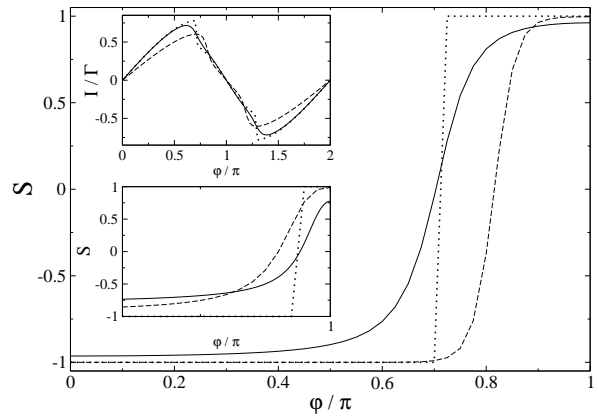}}
\caption{\label{f1}
Conformational state $S(\varphi)=S(-\varphi)$ 
vs superconductor's phase difference $\varphi$.
Results from Eq.~\eqref{conf} for $\Delta\gg \Gamma$ are shown for 
tunnel amplitudes $W_0 = 0$ (dotted) and $W_0=0.04\Gamma$ (solid),
with $\lambda = \epsilon = \Gamma/2$  and
$E_0 = 0.14 \Gamma$.  The dashed curve gives the exact result 
 for $W_0 = 0$  and $\Delta = 5\Gamma$, see Eq.~(\ref{andre}),  
extended to finite temperature $T = 0.01\Gamma$.  The upper inset shows the corresponding Josephson current-phase relations.  
Lower inset: Same as main figure but for $\Gamma = 4 \Delta$
with $\lambda = 2 \epsilon = \Delta/2$ and 
$E_0 = 0.45\Delta$. The dotted (solid) curve is obtained from 
the $\Gamma\gg \Delta$ effective Hamiltonian \eqref{heff2} 
with $W_0=0$ ($W_0 = 0.04\Delta$). The exact result
 for $W_0 = 0$ is shown as dashed curve for $T = 0.01\Delta$.  }
\end{figure}

Let us first illustrate our central findings when both the 
charging energy $U$ and the tunnel splitting $W_0$ are very small.  
Later on we show that for sufficiently small $U<U_c$, finite $U$ has no effect.
The ground-state energy $E_g = {\rm min} (E_+, E_-)$ then follows from the energies 
$E_\sigma=\sigma (\lambda-E_0)/2  - \epsilon_\sigma^A(\varphi)$
for fixed conformational state $\sigma=\pm$ with dot level
$\epsilon_\sigma=\epsilon+\sigma\lambda/2$. With Eq.~\eqref{sigma}
and $\tau_z={\rm diag}(1,-1)$,
the {\sl Andreev state} energy for arbitrary $\Delta/\Gamma$ follows from
\begin{equation}\label{andre}
\epsilon^A_\sigma(\varphi)=\epsilon^A_\sigma(0)+ \int\frac{d\omega}{2\pi} \ln
\frac{ \det \left 
[ i\omega-\tau_z\epsilon_\sigma -\Sigma(\omega;\varphi) \right ] }{ \det \left
[ i\omega-\tau_z\epsilon_\sigma -\Sigma(\omega;0) \right ] } .
\end{equation}
In the limits $\Gamma\gg \Delta$ and $\Delta\gg \Gamma$, this yields 
\cite{golubov}
\begin{eqnarray}\label{andreev}
\epsilon_\sigma^A(\varphi) &=& \Delta_\sigma \sqrt{1-{\cal T}_\sigma 
\sin^2(\varphi/2)}, \\ \nonumber
\Delta_\sigma&=& \left \{ \begin{array}{ll}  \frac{\Delta}{1+\Delta/\Gamma}, 
& \Gamma\gg \Delta,\\
\frac{\Gamma}{\sqrt{{\cal T}_\sigma}}, & \Delta\gg\Gamma, \end{array}  \right.
\end{eqnarray} 
with the normal transmission probability
${\cal T}_\sigma  =[1 +\epsilon_\sigma^2/\Gamma^2]^{-1}$.
As long as $E_+<E_-$ ($E_-<E_+$), we have $S(\varphi)=+1 (-1)$, i.e.~the
conformational state $\sigma=+ (-)$ is realized, 
with ideal (perfect) switching when the bands 
$E_+(\varphi)$ and $E_-(\varphi)$ cross at some phase $0<\varphi^*<\pi$.
Hence a necessary condition for switching follows: one of the two 
inequality chains (with ${\cal R}_\sigma=1-{\cal T}_\sigma$)
\begin{equation}\label{cond}
\Delta_+ \sqrt{{\cal R}_+} - \Delta_- \sqrt{{\cal R}_-} \lessgtr 
 \lambda-E_0 \lessgtr \Delta_+ - \Delta_-
\end{equation}
must be obeyed. If the dot level is close to a
resonance, $\epsilon_+ \approx 0$ or $\epsilon_-\approx 0$, the 
reflection probabilities ${\cal R}_+$ and ${\cal R}_-$ are 
significantly different, and Eq.~\eqref{cond} holds over a wide 
parameter range.  Then Eq.~\eqref{current} yields
\begin{equation}\label{jos1}
I(\varphi) = \frac{e \Delta_{S(\varphi)} }{2\hbar} 
\frac{{\cal T}_{S(\varphi)} \sin(\varphi)}{\sqrt{1-{\cal T}_{S(\varphi)}
\sin^2(\varphi/2)}}.
\end{equation}
In the regime \eqref{cond}, the transmission amplitude switches 
between  ${\cal T}_+$ and ${\cal T}_-$ when $\varphi=\varphi^*$.  
This implies non-standard current-phase relations, as shown in the upper
inset of Fig.~\ref{f1}.

Having established the basic phenomenon,  we now address the effects of 
finite $U$ and/or tunneling $W_0$.  Progress can be made in the 
limits $\Gamma\gg \Delta$ and $\Delta\gg \Gamma$.
Let us start with the case when $\Delta$ is the largest 
energy scale of relevance. Then the dynamics
is always confined to the subgap regime (Andreev states), 
and quasiparticle tunneling processes from the leads (continuum states)
are negligible. Technically, Eq.~\eqref{sigma} can then be 
replaced by $\Sigma(\tau)=\Gamma \cos(\varphi/2) \delta(\tau) \left(
\begin{array}{cc} 0 & 1 \\ 1 & 0 \end{array}\right)$,
and the problem is equivalently described by the effective Hamiltonian
$H_{\rm eff} = H_0 + \Gamma \cos(\varphi/2) \left( d_\downarrow d_\uparrow +
 d_\uparrow^\dagger d_\downarrow^\dagger \right)$.
The resulting Hilbert space can be decomposed into 
orthogonal subspaces, ${\cal H}={\cal H}_A 
\otimes {\cal H}_S$, where the Andreev sector ${\cal H}_A$ is 
spanned by the zero- and two-electron dot states $|0\rangle$ and
$|2\rangle = d_\uparrow^\dagger d^\dagger_\downarrow |0\rangle$ (and, of course, by the conformational TLS states), while ${\cal H}_S$ is spanned by 
the one-electron states $|s\rangle \equiv d_s^\dagger |0\rangle$. 
For convenience shifting  $H_{\rm eff}\to H_{\rm eff}-\epsilon$, the 
single-particle sector has a pair of doubly-degenerate 
eigenenergies $-\frac{U}{2} \pm \epsilon_S$ with
$\epsilon_S = \frac12 \sqrt{(E_0-\lambda)^2 + W_0^2}$, whereas
the Andreev sector is described by  
\begin{equation} \label{heff}
H_{\rm eff}^A = \frac{\lambda-E_0}{2} \sigma_z 
-\frac{W_0}{2} \sigma_x + \frac{\lambda}{2} \tau_z \sigma_z
+ \epsilon\tau_z + \Gamma\cos(\varphi/2) \tau_x
\end{equation} 
with Pauli matrices
$\tau_{x,z}$ acting in $\{ |2\rangle, |0\rangle\}$ subspace.
If the ground state of $H_{\rm eff}$ lies in the Andreev sector, 
the Josephson current can be non-zero, while otherwise $I=0$ due
to the $\varphi$-independence of the single-particle sector.  
For sufficiently strong interactions, $U>U_c(\varphi)$, 
the ground state of $H_{\rm eff}$ is
in the single-particle sector ${\cal H}_S$.
This is indicative of a quantum phase transition to the
magnetic $\pi$-junction regime \cite{bcs-anderson}. While
this regime is outside the scope of Eq.~\eqref{heff} 
(since continuum states are not included),
we have confirmed this scenario by a perturbative 
calculation expanding in $\Gamma$ for the full model.
For $\lambda\to 0$, we find $U_c=2 \sqrt{\epsilon^2+
 \Gamma^2\cos^2(\varphi/2)}$, see Eq.~\eqref{andreev} for $\Delta\gg \Gamma$. 
Note that $\epsilon$ and hence $U_c$ can in principle 
be tuned by a gate voltage.  For
$\lambda\gg {\rm max}( |E_0|,|\epsilon|,\Gamma)$, 
we instead find $U_c=\lambda$.  Because $H_{\rm eff}^A$ is independent 
of $U$ (up to the shift $\epsilon=\epsilon_d+U/2$), 
a weak interaction $U<U_c$ has no 
effect, and in what follows we set $U=0$.  Since a coupling of the 
TLS to the dot's {\sl spin} involves only the $\varphi$-independent 
subspace ${\cal H}_S$, such couplings are also of little
relevance for switching, in accordance with the 
small polar displacements predicted for spins in a 
Josephson junction \cite{molmag}.

Physical observables can then be computed from $H_{\rm eff}^A$ in
Eq.~\eqref{heff}. The eigenenergies are roots to the 
exactly solvable quartic equation 
\begin{equation}\label{quartic}
E^4 -2 \Lambda_2 E^2 + \Lambda_1 E + \Lambda_0 = 0,
\end{equation}
with coefficients $\Lambda_2 = \epsilon_A^2+\epsilon_S^2+\lambda^2/4,
\Lambda_1 = 2\lambda\epsilon(E_0- \lambda)$ and 
$\Lambda_0 = \left(\epsilon^2_A-\epsilon_S^2+\lambda^2/4\right)^2
-\lambda^2 (\epsilon^2-W_0^2/4).$ The lowest-lying of the four roots 
yields the exact but lengthy result for the ground-state energy $E_g$.
Convenient expressions for $S(\varphi)$ and $I(\varphi)$ 
in Eq.~\eqref{current} follow by taking the respective derivatives directly in
 Eq.~\eqref{quartic}. For instance, with 
$\Lambda_i^\prime=\partial \Lambda_i/\partial E_0$,
the conformational variable reads
\begin{equation}\label{conf}
S(\varphi) = - \frac{2\Lambda_2^\prime E_g^2- \Lambda_1^\prime E_g 
- \Lambda_0^\prime}{2E_g (E_g^2 - \Lambda_2)+\Lambda_1/2}.
\end{equation}
Typical results for $S(\varphi)$ and $I(\varphi)$ are shown in Fig.~\ref{f1}.  
The most efficient way to induce conformational changes, including a complete
(symmetric) reversal $S\to -S$, is achieved in the weak-coupling regime
 $\lambda\ll \epsilon_A$, 
where the four roots to Eq.~\eqref{quartic} can be simplified to
\begin{equation}\label{adiabatic}
E_{\pm,\pm}=\pm \epsilon_A(\varphi) \pm \frac12 \sqrt{W_0^2 + [\lambda(1-
\epsilon/\epsilon_A(\varphi))-E_0]^2},
\end{equation}
with ground-state energy $E_g=E_{--}$.  Remarkably, Eq.~\eqref{adiabatic} 
remains accurate even for $\lambda \approx \epsilon_A$.
A complete reversal is achieved when tuning $E_0$ or $\epsilon$ such that 
$E_0=\lambda[1-{\rm sgn}(\epsilon)] -{\cal F}$ with ${\cal F}=-
 \frac{\lambda}{2} {\rm sgn}(\epsilon) \left[ 1-|\epsilon|/\epsilon_A(0)
\right]$. In that case, $S(0)=-S(\pi)={\cal F}/\sqrt{W_0^2+{\cal F}^2}.$
 When comparing to the $W_0=0$ result,
we observe that a finite tunnel amplitude $W_0$ only leads to a rounding
of the transition and a decrease in the switching amplitude, 
but it does not destroy the effect.  Finally, with Eq.~\eqref{adiabatic}, 
the Josephson current in the weak-coupling limit is
\begin{equation}
I(\varphi) = \frac{\Gamma^2 \sin\varphi}{2\epsilon_A(\varphi)}
\left( 1+ \frac{\lambda\epsilon}{2 \epsilon_A^2(\varphi)} S(\varphi) \right).
\end{equation}

Next we briefly discuss the opposite limit within a similar 
truncation scheme, setting $U=0$. For $\Gamma\gg \Delta$ and 
$\varphi\neq 2\pi n$ (integer $n$), 
the relevant subgap dynamics is again captured by an effective
two-level Hamiltonian describing the Andreev states \cite{andreev2}, 
coupled to the conformational TLS.  With 
Pauli matrices $\tau_{x,y,z}$ in Andreev level subspace and 
the notation [see Eq.~\eqref{andreev}]
\[
H_\sigma = \Delta_\sigma e^{-i\tau_y \sqrt{{\cal R}_\sigma}\varphi/2} 
 \left( 
\sqrt{{\cal R}_\sigma} \ \sin(\varphi/2) \tau_z + 
\cos(\varphi/2) \tau_x \right),
\]
 the effective Hamiltonian follows in TLS space as 
\begin{equation}\label{heff2}
H_{\rm eff} =\left(\begin{array}{cc}
\frac{\lambda-E_0}{2}+ H_+ & -\frac{W_0}{2}  \\
  -\frac{W_0}{2} & -\frac{\lambda-E_0}{2}+ H_-\end{array}\right).
\end{equation}
Physical observables are then easily obtained, see
the lower inset of Fig.~\ref{f1}.
Again the qualitative features of the $W_0=0$ solution persist. 

\begin{figure}
\scalebox{0.32}{\includegraphics{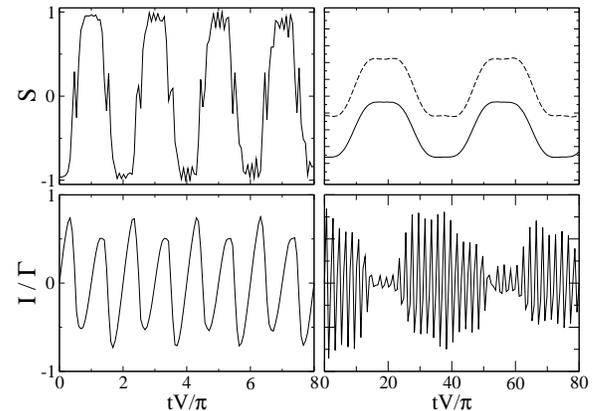}}
\caption{ \label{f2}
Time dependence of $S$ (upper row) and ac Josephson current $I$ 
(lower row).  The left panel shows the 
adiabatic evolution for low voltage, $V = 0.01\Gamma$.
Parameters are the same as for the solid curve in
the main panel of Fig.~\ref{f1}.  The right panel is for $V = 5\Gamma$ with 
$\epsilon/\Gamma = 0.2$ (solid) and 0.6 (dashed, current not shown).
Other parameters are $\lambda = \Gamma/2$ and $E_0 = W_0 = 0.2\Gamma$.
}
\end{figure}

The effective Hamiltonian \eqref{heff} for $\Delta\gg \Gamma$ 
also allows to study the voltage-biased junction  with
$V\ll \Delta$, where the superconducting phase difference is
time-dependent, $\varphi(t)=2V t$.  During the time evolution induced
by $\varphi(t)$, the Andreev and single-particle Hilbert subspaces 
${\cal H}_A$ and ${\cal H}_S$ remain decoupled and mutually orthogonal. 
The task is therefore reduced to solving the time-dependent Schr\"odinger
equation $i\partial_t \Psi(t) = H_{\rm eff}^A(t) \Psi(t),$
where $\Psi(t)$ is a 4-component wave function representing the 
two Andreev states and the TLS, and $H_{\rm eff}^A(t)$
is given by Eq.~\eqref{heff} with $\varphi\to 2Vt$. 
For this description to hold at finite $\Delta$, the escape rate $\gamma$ 
of Andreev state quasiparticles into the continuum states of the leads should be
negligibly small.  The rate $\gamma$ follows from 
the tunneling self-energy, see Ref.~\cite{alfredo}
for the opposite limit $\Gamma\gg \Delta$.  For $\epsilon=0$, we find 
\begin{equation}
\gamma \simeq \Gamma \exp\left( -\frac{2\Delta}{V} \left[\ln(2\Delta/\Gamma)-1
\right]\right),
\end{equation}
leading to exponentially small rates for realistic system parameters 
throughout the regime $\Delta\gg \Gamma$.
Numerical solution of the time-dependent Schr\"odinger equation 
leads to the results in Fig.~\ref{f2}. 
 They can be understood in terms of the 
four eigenenergies \eqref{adiabatic}.
For small $V$, the time evolution is 
basically adiabatic, and the LZ probability is very small.
The left panel in Fig.~\ref{f2} shows such an adiabatic evolution involving
time-periodic level crossings of the bands $E_{--}$ and $E_{-+}$ in 
Eq.~\eqref{adiabatic}, thereby explaining the existence of two
different supercurrent oscillation amplitudes.  The ``noisy'' features in $S(t)$ are 
fully reproducible and reflect a superposition of almost filled 
and almost empty levels.  There are no LZ transitions in that limit, but only a
continuous change of energy bands at the branching times where $E_{--}=E_{-+}$. 
However, for larger $V/\Gamma$, the LZ probability becomes sizeable
and the dynamics is more complex, generally involving a dynamical 
population of all four subgap states.  The right panel in Fig.~\ref{f2}
displays the case of relatively large $V$, where the system oscillates
due to LZ transitions between the levels $E_{--}$ and $E_{+-}$. 
The frequency $\omega_S$ of the $S(t)$ oscillations is much slower than the 
 Josephson frequency $\omega_J=2eV/\hbar$ and determined by the
lowest interlevel transition energy, $\omega_S={\rm min}(E_{-+}-E_{--})$. 
Note that $\omega_S$ reappears in the ac Josephson current.
 
To conclude, we predict that the conformational degree of freedom 
(represented by a TLS) in a superconducting molecular 
dot or break junction responds in a dissipationless manner
to variations of the phase difference $\varphi$ across the 
dot/junction, including a complete reversal.  This effect should be 
observable using existing experimental methods over a wide parameter
range. Under an applied voltage, this effect leads to quasi-periodic 
TLS dynamics due to the time-dependent occupation
probabilities of Andreev states.--- We thank T. Martin for discussions.
This work was supported by the SFB TR 12 of the DFG and by the
EU networks INSTANS and HYSWITCH.


\begin{thebibliography}{99}
\bibitem{foot0}
Consider the dot in Eq.~\eqref{h0} capacitively side-coupled to a 
double dot (fermion operators $f_{1,2}$) tuned to 
single occupancy, $f^\dagger_1 f^{}_1 +f^\dagger_2 f^{}_2=1$.
Then the TLS-plus-coupling part of $H_0$ is 
$-\frac{W_0}{2}(f^\dagger_1 f^{}_2 + f_2^\dagger f_1^{}) - \frac{E_0}{2}
(f^\dagger_1 f^{}_1 - f^\dagger_2 f^{}_2) + \sum_{i=1,2} U_i f_i^\dagger
f^{}_i (n_\uparrow + n_\downarrow)$, and $\lambda$ 
corresponds to the difference in capacitive couplings $U_1-U_2$.
\bibitem{foot}
The generalization to asymmetric cases is straightforward and 
does not yield new physics.
\end{thebibliography}
\end{document}